\documentclass[twoside,twocolumn,english,superscriptaddress,showpacs]{revtex4-2}
\usepackage[T1]{fontenc}
\usepackage[utf8]{inputenc}
\usepackage{babel}
\usepackage{bm}
\usepackage{amsmath}
\usepackage[unicode=true,pdfusetitle,
 bookmarks=true,bookmarksnumbered=true,bookmarksopen=true,bookmarksopenlevel=1,
 breaklinks=false,pdfborder={0 0 0},pdfborderstyle={},backref=false,colorlinks=false]
 {hyperref}

\makeatletter
\usepackage{braket}
\usepackage{bm}
\usepackage{tikz}
\usepackage{pgfplots}
\usepackage{pgf}
\usepackage{subfigure}
\usepackage{nicematrix}
\usepackage{diagbox}

\makeatother

\begin{document}
\title{Quantum metric induced hole dispersion and emergent particle-hole symmetry in topological flat bands}
\author{Guangyue Ji}
\affiliation{Division of Physics and Applied Physics, Nanyang Technological University,
Singapore 637371}
\author{Bo Yang }
\affiliation{Division of Physics and Applied Physics, Nanyang Technological University,
Singapore 637371}
\begin{abstract}
The emergent hole dispersion in flat bands is an invaluable platform to study the interplay of quantum geometry and electron-electron interaction with a relatively simple setting. In this work, we find that the hole dispersion in ideal bands has a linear relationship with the trace of the quantum geometry tensor at every $\bm{k}$-point for a wide range of interactions to a good approximation. Next, we give a microscopic analysis on the hole dispersion and show that the linear relationships for short-range and long-range interactions in $\bm{k}$-space have different origins. Moreover, we show how to exploit this observation to engineer particle-hole symmetry in a Chern band with fluctuating quantum geometry. Our results will be useful for further studying the physics in particle-hole symmetric flat bands both in theory and in experiment.
\end{abstract}
\maketitle

\textit{Introduction.---}
Recent theoretical and experimental developments in flat Chern bands reveal fascinating physics in both single-particle and strongly-correlated topological phases at zero magnetic field$\,$\citep{Wu2019TMD,devakul2021magic,cai_signatures_2023,zeng_thermodynamic_2023,park_observation_2023,lu_fractional_2024,goldman_zero-field_2023,dong_composite_2023,wang_fractional_2024,Reddy_FQAHS_2023,Reddy_2023_Toward,Morales2024magic}. Among various ingredients, the quantum geometry tensor (QGT) $\mathcal{Q}_{\bm{k}}$ (i.e., the Fubini-Study metric $g_{\bm{k}}$ and the Berry curvature $\Omega_{\bm{k}}$) of flat bands plays a crucial role in our understanding of the underlying physics. At the single-particle level, the quantum metric tensor can characterize the size of a wave packet in a flat band, e.g., the coherence length of a flatband superconductor is dominantly controlled by the quantum metric$\,$\citep{tian2023evidence,chen_ginzburg-landau_2024}, and the Berry curvature governs various intrinsic anomalous transport$\,$\citep{xiao_berry_2010,Nagaosa2010Anomalous}. Moreover, at the many-body level, the interplay of QGT and electron-electron interaction can lead to much richer physics. In particular, recent works have focused on studying the interaction dynamics of flat bands mimicking the QGT of Landau levels (LLs), such as ideal bands, vortexable bands, generalized LLs and so on$\,$\citep{LIU2024515,wang_exact_2021,wang_hierarchy_2022,wang_origin_2023,ledwith_vortexability_2023,fujimoto_higher_2024,liu2024theory,wang2024higher,ahn_first_2024}. However, due to a lack of a unified paradigm (e.g., Haldane pseudopotentials for LLs$\,$\citep{haldane_fractional_1983}) for analyzing the complex interaction effects, the connection between various energy spectrums and QGT is still obscure. 

Fortunately, the emergent hole dispersion in a flat band provides us an invaluable platform to study the interplay of QGT and electron-electron interaction with a relatively simple setting. A single hole is created by removing an electron from a fully filled band. In the absence of particle-hole (PH) symmetry, the hole acquires an emergent dispersion due to the interaction between electrons, independent of the electron kinetic energy$\,$\citep{Grushin2012Enhancing,lauchli_hierarchy_2013,abouelkomsan_particle-hole_2020,abouelkomsan_quantum_2023,liu2024broken}. Its dynamics is completely determined by interactions yet still closely related to the single-particle QGT. For interactions only allowing small momentum transfer limit, the connection between the hole dispersion and the QGT has been revealed in Ref.$\,$\citep{abouelkomsan_quantum_2023}. Moreover, the hole dispersion will play an important role on interacting phases especially at the high fractional fillings, which are relevant to the PH conjugate of the familiar fractional quantum Hall phases and compressible composite fermion liquids$\,$\citep{lauchli_hierarchy_2013,abouelkomsan_particle-hole_2020,abouelkomsan_quantum_2023,Reddy_FQAHS_2023,Reddy_2023_Toward,liu2024broken}.

In this letter, we study the emergent hole dispersion in flat bands and reveal rather interesting behaviours that are much more universal than what was recently discussed in Ref.$\,$\citep{abouelkomsan_quantum_2023}. We find that the hole dispersion in ideal bands has a linear relationship with the trace of QGT at every $\bm{k}$-point for a wide range of interactions to a good approximation. While this is expected for interactions with only very short momentum transfer (i.e., very long range in real space), we propose an analytic approach to partially explain this could also be the case for the realistic short-range interactions in real space, including the Coulomb and pseudopotential interactions. This rather surprising result is useful for the engineering of a family of interactions which render the Chern band PH symmetric, a property present in the LLs but is generally missing for any Chern bands with non-uniform QGT in the momentum space. We also discuss how to engineer this emergent PH symmetry in real materials, e.g., the twisted bilayer $\textrm{WSe}_{2}$ ($t\textrm{WSe}_{2}$). Further research on the physics of particle-hole symmetric flat bands can be conducted based on our results.

\textit{Microscopic model for hole dispersion.---}
The interaction Hamiltonian projected a flat band takes the following general form 
\begin{equation}
\hat{H}=\sum_{\bm{k}_{1}\bm{k}_{2}\bm{k}_{3}\bm{k}_{4}}^{\prime}V_{\bm{k}_{1}\bm{k}_{2}\bm{k}_{3}\bm{k}_{4}}\hat{c}_{\bm{k}_{1}}^{\dagger}\hat{c}_{\bm{k}_{2}}^{\dagger}\hat{c}_{\bm{k}_{3}}\hat{c}_{\bm{k}_{4}},\label{eq:Interaction_Hamiltonian}
\end{equation}
where $\hat{c}_{\bm{k}}^{\dagger}$ ($\hat{c}_{\bm{k}}$) is the creation (annihilation) operator of the Bloch eigenstate $\vert\psi_{\bm{k}}\rangle$ in the flat band with momentum $\bm{k}$, the summation of the momentum with prime is within the first Brillouin zone (1BZ). The explicit expression of the interaction matrix element is given by
\begin{align}
V_{\bm{k}_{1}\bm{k}_{2}\bm{k}_{3}\bm{k}_{4}} & =\sum_{\bm{b},\bm{b}^{\prime}}\delta_{\bm{k}_{1}+\bm{k}_{2}-\bm{k}_{3}-\bm{k}_{4}+\bm{b}-\bm{b}^{\prime},0}v_{\bm{k}_{1}-\bm{k}_{4}+\bm{b}}\nonumber \\
 & \times\phi_{\bm{k}_{4},-\bm{b}}\phi_{\bm{k}_{3},\bm{b}^{\prime}}\langle u_{\bm{k}_{1}}\vert u_{\bm{k}_{4}-\bm{b}}\rangle\langle u_{\bm{k}_{2}}\vert u_{\bm{k}_{3}+\bm{b}^{\prime}}\rangle,\label{eq:Interaction_matrix}
\end{align}
where $\bm{b}$ and $\bm{b}^{\prime}$ are reciprocal wave vectors, the delta function comes from the momentum conversation for a translation-invariant interaction, $v_{\bm{q}}=\int d^{2}\bm{r}\text{\,}v(\bm{r})e^{-i\bm{q}\cdot\bm{r}}$
is the Fourier transform of the real-space interaction $v(\bm{r})$, $\vert u_{\bm{k}}\rangle=e^{-i\bm{k}\cdot\hat{\bm{r}}}\vert\psi_{\bm{k}}\rangle$ is the periodic part of the Bloch state $\vert\psi_{\bm{k}}\rangle$, $\phi_{\bm{k},\bm{b}}$ is a phase factor resulting from the boundary condition $\phi_{\bm{k},\bm{b}}\hat{c}_{\bm{k}}=\hat{c}_{\bm{k}+\bm{b}}$ in $\bm{k}$-space. 

A single hole is created by removing an electron from a fully filled flat band. Different from a single electron in the flat band, of which the kinetic energy is zero and has no dispersion, the hole can have an emergent dispersion proportional to the interaction energy scale that in some materials is dominant over not only the bandwidth but also the band gap$\,$\citep{devakul2021magic,Reddy_FQAHS_2023,kourtis_fractional_2014}. If we see the fully filled band as the vacuum state with zero-energy, the hole state has an energy $\varepsilon_{\bm{k}}=\sum_{\bm{k}^{\prime}}^{\prime}\left(V_{\bm{k}^{\prime}\bm{k}\bm{k}^{\prime}\bm{k}}+V_{\bm{k}\bm{k}^{\prime}\bm{k}\bm{k}^{\prime}}-V_{\bm{k}\bm{k}^{\prime}\bm{k}^{\prime}\bm{k}}-V_{\bm{k}^{\prime}\bm{k}\bm{k}\bm{k}^{\prime}}\right)$, which is the summation of the interaction energy of this electron with all other electrons$\,$\citep{lauchli_hierarchy_2013}. For an inversion-invariant interaction, it can be simplified to be
\begin{align}
\varepsilon_{\bm{k}} & =2(\varepsilon_{F,\bm{k}}+\varepsilon_{H,\bm{k}})=2\sum_{\bm{k}^{\prime}}^{\prime}\left(V_{\bm{k}^{\prime}\bm{k}\bm{k}^{\prime}\bm{k}}-V_{\bm{k}\bm{k}^{\prime}\bm{k}^{\prime}\bm{k}}\right),\label{eq:Hole_dispersion}
\end{align}
where $\varepsilon_{F,\bm{k}}$ and $\varepsilon_{H,\bm{k}}$ are the Fock and Hartree interaction energies, respectively. By substituting Eq.$\,$(\ref{eq:Interaction_matrix}) into the above equation, we can obtain the explicit expressions of $\varepsilon_{F,\bm{k}}$ and $\varepsilon_{H,\bm{k}}$ and they are
\begin{align}
\varepsilon_{F,\bm{k}} & =\sum_{\bm{q}}v_{\bm{q}}\vert\langle u_{\bm{k}}\vert u_{\bm{k}+\bm{q}}\rangle\vert^{2},\label{eq:First_term}\\
\varepsilon_{H,\bm{k}} & =-\sum_{\bm{b}}v_{\bm{b}}C_{\bm{b}}\phi_{\bm{k},-\bm{b}}\langle u_{\bm{k}}\vert u_{\bm{k}-\bm{b}}\rangle,\label{eq:Second_term}
\end{align}
where the summation of the momentum $\bm{q}$ without prime is for all space and $C_{\bm{b}}=\sum_{\bm{k}^{\prime}}^{\prime}\phi_{\bm{k}^{\prime},\bm{b}}\langle u_{\bm{k}^{\prime}}\vert u_{\bm{k}^{\prime}+\bm{b}}\rangle$. 

\begin{figure}[t] 
\includegraphics[width=0.98\columnwidth]{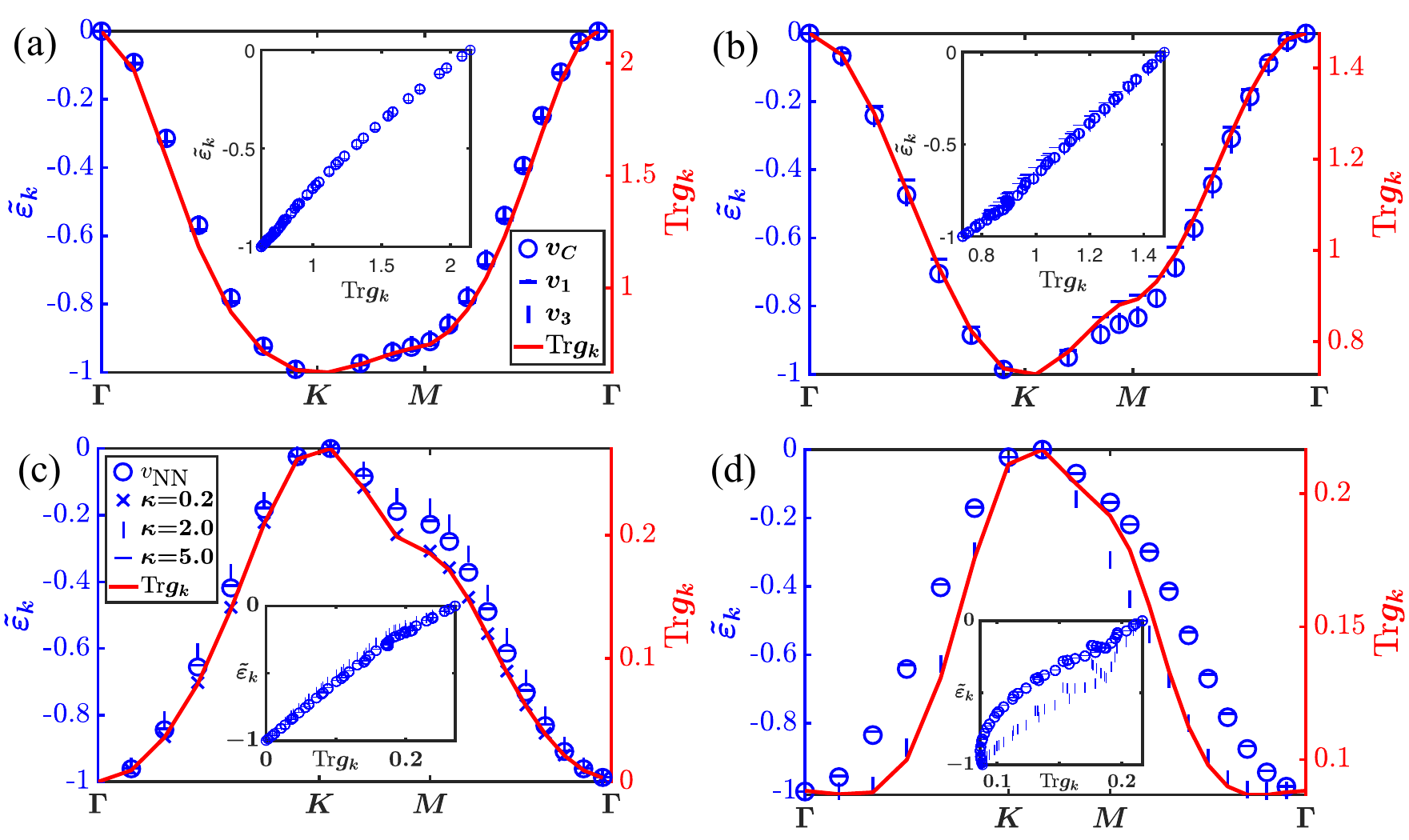}
\caption{\label{fig: Hole_energy_vs_QGT}
(a) A single-hole dispersion in cTBG at the first magnetic angle for different kinds of interactions, including the Coulomb interaction $v_{C}$ (blue circles), Haldane pseudopotentials $v_{1}$ (horizontal lines) and $v_{3}$ (vertical lines). The normalized hole dispersion is defined as $\tilde{\varepsilon}_{\bm{k}}{=}{-}(\varepsilon_{\bm{k}}{-}\min\epsilon_{\bm{k}})/(\max\varepsilon_{\bm{k}}{-}\min\varepsilon_{\bm{k}})$. It has a well linear dependence on the trace of metric tensor $\text{Tr}g(\bm{k})$ for these three interactions as shown in the inset. (b) is similar to (a) except at the second magnetic angle of cTBG.
(c) Hole dispersion in the lowest flat band of kagome lattice model for different kinds of interactions, including the nearest-neighbor interaction $v_{\text{NN}}$ (blue circles) and Gaussian interaction $v_{G}{=}\exp(-\vert\bm{q}\vert^{2}/\kappa^{2})$ with different decay lengths $\kappa{=}0.2, 2, 5$ (blue crosses, vertical lines and horizonal lines, respectively).(d) is similar to (a) except for the $t\textrm{WSe}_{2}$. The unit length for ideal bands, lattice models and $t\textrm{WSe}_{2}$ are the magnteic length, lattice constant and moir\'e superlattice constant, respectively.
} 
\end{figure}

\textit{Universal quantum metric induced hole dispersion.---}
Even though the hole dispersion completely origins from electron-electron interactions, here we show in some cases it is almost entirely controlled by the single-particle QGT. This is highly nontrivial because the QGT is only the long wavelength part of the band form factor, and for a generic interaction the form factor at all wavelengths should contribute. In particular for ideal bands, the hole dispersion has a linear relationship with the trace of metric tensor at every $\bm{k}$-point for a wide range of interactions to a good approximation as shown in Fig.$\,$\ref{fig: Hole_energy_vs_QGT} (a) and (b). Ideal bands are bands which exactly satisfy the trace condition $\text{Tr}g_{\bm{k}}=\Omega_{\bm{k}}$. Their wave functions take the general form of $\psi_{\bm{k}}(\bm{r})=\mathcal{N}_{\bm{k}}\mathcal{B}(\bm{r})\Phi_{\bm{k}}(\bm{r}),$ where $\mathcal{N}_{\bm{k}}$ is the normalization coefficient, $\mathcal{B}(\bm{r})$ is a real-space function which depends on the crystal structure, and $\Phi_{\bm{k}}(\bm{r})$ is the LLL wave function$\,$\citep{wang_exact_2021}. The hole dispersion in ideal bands is given by
\begin{multline}
\varepsilon_{\bm{k}}^{\text{IB}}=2\sum_{\bm{k}^{\prime}}^{\prime}\sum_{\bm{b},\bm{b}^{\prime},\bm{b}^{\prime\prime}}\mathcal{N}_{\bm{k}}^{2}\mathcal{N}_{\bm{k}^{\prime}}^{2}w_{\bm{b}^{\prime}}w_{\bm{b}^{\prime\prime}}\\
\times\left[v_{\bm{k}^{\prime}-\bm{k}-\bm{b}}f_{\bm{b}-\bm{b}^{\prime}}^{\bm{k}^{\prime}\bm{k}}f_{-\bm{b}-\bm{b}^{\prime\prime}}^{\bm{k}\bm{k}^{\prime}}-v_{-\bm{b}}f_{\bm{b}-\bm{b}^{\prime}}^{\bm{k}\bm{k}}f_{-\bm{b}-\bm{b}^{\prime\prime}}^{\bm{k}^{\prime}\bm{k}^{\prime}}\right].\label{eq:Ideal_band_dispersion}
\end{multline}
where $w_{\bm{b}}$ is the Fourier component of $\vert\mathcal{B}(\bm{r})\vert^{2}$, $f_{\bm{b}}^{\bm{k}\bm{k}^{\prime}}=\eta_{\bm{b}}e^{\frac{i}{2}(\bm{k}+\bm{k}^{\prime})\times\bm{b}}e^{\frac{i}{2}\bm{k}\times\bm{k}^{\prime}}e^{-\frac{1}{4}\vert\bm{k}-\bm{k}^{\prime}-\bm{b}\vert^{2}}$, $\eta_{\bm{b}}=(-1)^{mn+m+n}$ for $\bm{b}=m\bm{b}_{1}+n\bm{b}_{2}$ and $\bm{b}_{i}$ are primitive reciprocal wave vectors. Fig.$\,$\ref{fig: Hole_energy_vs_QGT} (a) and (b) show the hole dispersion of chiral twisted bilayer graphene (cTBG) at the first and second magic angles respectively, which are typical examples of ideal bands with fluctuated QGT. For a wide range of interactions, including the Coulomb interaction $v_{C}=1/\vert\bm{q}\vert$, the Haldane pseudopentials$\,$\citep{haldane_fractional_1983} $v_{1}=1-\vert\bm{q}\vert^{2}$ and $v_{3}=1-3\vert\bm{q}\vert^{2}+\frac{3}{2}\vert\bm{q}\vert^{4}-\frac{1}{6}\vert\bm{q}\vert^{6}$, the linear relationship between $\varepsilon_{\bm{k}}^{\text{IB}}$ and $\text{Tr}g(\bm{k})$ is fairly accurate. Since $\text{Tr}g_{\bm{k}}=\Omega_{\bm{k}}$ for ideal bands, the linear relationship also establishes between the hole dispersion and the Berry curvature.

This phenomenon is rather general, in particular for other flat Chern bands approximately satisfying the trace condition, though the linear relationship approximation in these systems is not as robust as ideal bands. For example, we can look at two typical systems, i.e., the kagome lattice model$\,$\citep{tang_high-temperature_2011} and $t\textrm{WSe}_{2}$ $\,$\citep{devakul2021magic}. The hole dispersion of lattice systems takes the following general form
\begin{multline}
\varepsilon_{\bm{k}}^{\text{lat}}=2\sum_{\bm{k}^{\prime}}^{\prime}\sum_{\bm{b}}\sum_{a,b}e^{i\bm{b}\cdot(\bm{\delta}_{a}-\bm{\delta}_{b})}\\
\times\left[v_{\bm{k}^{\prime}-\bm{k}+\bm{b}}\langle u_{\bm{k}^{\prime}}^{a}\vert u_{\bm{k}}^{a}\rangle\langle u_{\bm{k}}^{b}\vert u_{\bm{k}^{\prime}}^{b}\rangle-v_{\bm{b}}\right],
\end{multline}
where $a$, $b$ are sublattice indices, $\bm{\delta}_{a}$ is the relative position of the $a$ site from the unit cell position $\bm{R}_{i}$, the form factor $\langle u_{\bm{k}^{\prime}}^{a}\vert u_{\bm{k}}^{a}\rangle=\psi_{\bm{k}^{\prime}}^{a*}\psi_{\bm{k}}^{a}$ and $\psi_{\bm{k}}^{a}$ is the $a$-sublattice component of the eigenstate of the single-particle kinetic Hamiltonian.  Different from continuous systems, the form factor of lattice systems is periodic and the Hartree term has no contribution to the hole dispersion. Fig.$\,$\ref{fig: Hole_energy_vs_QGT} (c) shows the hole dispersion in the kagome model. Here, we consider the most common nearest-neighbor interaction $v_{\text{NN}}(\bm{r})$ and the tunable Gaussian interaction $v_{G}=\exp(-\vert\bm{q}\vert^{2}/\kappa^{2})$. For a small decay length $\kappa=0.2$, the linear relationship is pretty good. For $\bm{k}$-space longer-range Gaussian interactions and the nearest-neighbor interaction, the hole dispersion has a small deviation from the trace of metric tensor. For the continuous system $t\textrm{WSe}_{2}$, which deviates from the trace condition much more, the deviation from the linear relationship is also much larger than ideal bands. However, the monotonic relationship still looks not bad for a wide range of interactions. 

\textit{Microscopic analysis.---}
We will now carry out theoretical analysis to show why the phenomenon shown above is nontrivial, with some partial analytic explanations. In the $\bm{k}$-space short-range interaction limit\citep{abouelkomsan_quantum_2023}, the Fock energy $\varepsilon_{F,\bm{k}}\approx\sum_{\bm{q}}v_{\bm{q}}[1-g_{\mu\nu}q_{\mu}q_{\nu}+O(q^{3})]$, while the Hartree energy $\varepsilon_{H,\bm{k}}\approx-N_{s}v_{\bm{0}}$, which can be seen as a uniform background contribution and has no contribution to the hole dispersion. Thus, the hole dispersion in this limit completely comes from the Fock energy and has a linear dependence on the trace of metric tensor. However, the above argument doesn't apply for most common interactions. For example, the summation $\sum_{\bm{q}}v_{\bm{q}}q_{\mu}q_{\nu}$ and the summation of higher order terms for the Coulomb interaction $v_{\bm{q}}=1/\vert\bm{q}\vert$ are divergent; moreover the higher-order terms are generally not small and cannot be ignored. These terms are no longer explicitly related to the QGT. 

\begin{figure}[t] 
\includegraphics[width=0.98\columnwidth]{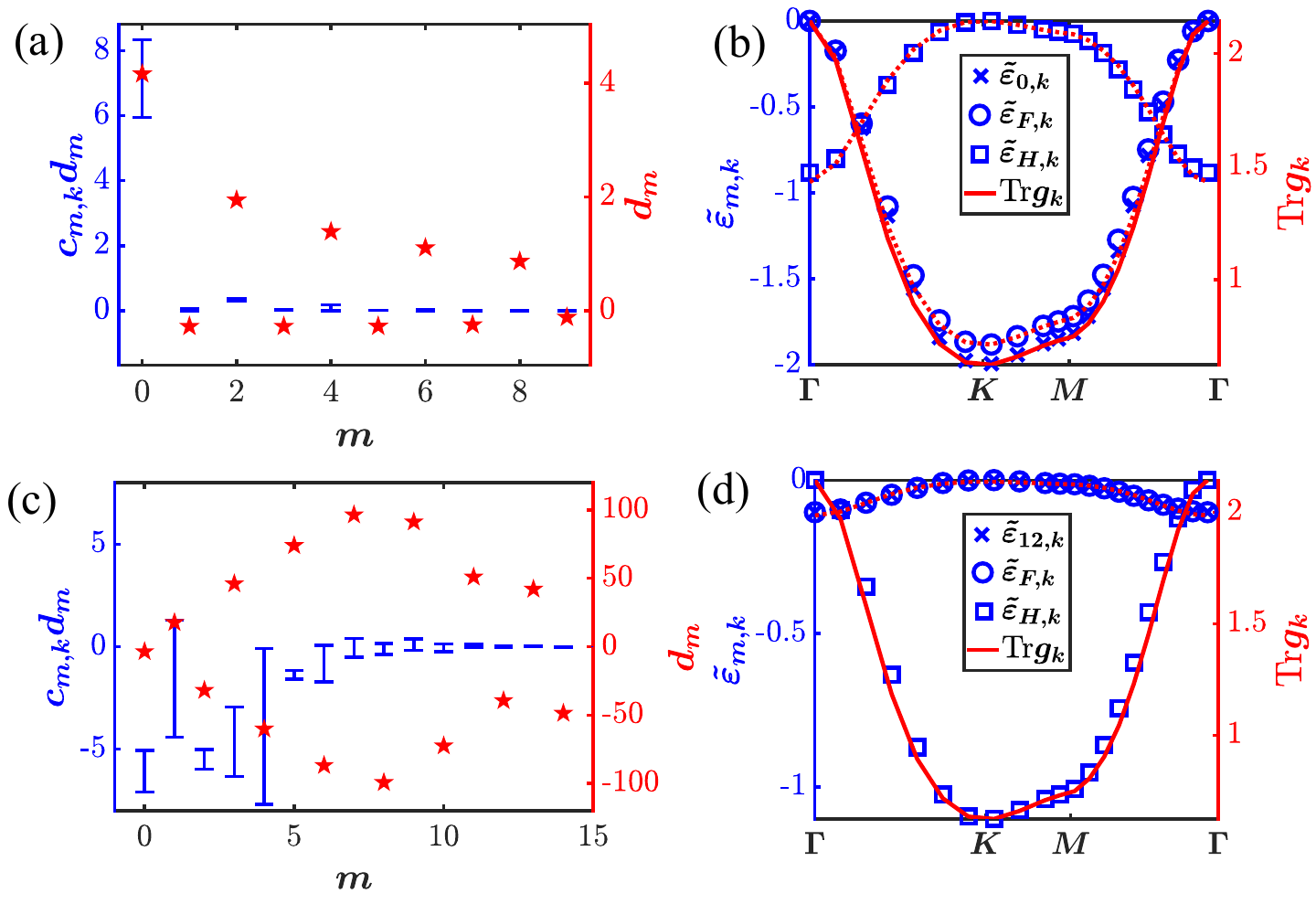}
\caption{\label{fig: Microscopic_analysis}
(a) The expansion coefficients $d_{m}$ (red pentagrams) and $c_{m,\bm{k}}d_{m}$ (blue bars) for cTBG at the first magnetic angle with Coulomb interaction. The orthonormal basis are Laguerre polynomials with the magnetic length $l_B$ as the characteristic length. The upper and  lower bars denote the maximum and minimum of  $c_{m,\bm{k}}d_{m}$ for a given $m$, respectively. 
(b)  The renormalized energy $\tilde{\varepsilon}_{m,\bm{k}}=-2(\varepsilon_{m,\bm{k}}-\min \varepsilon_{m,\bm{k}})/(\max \varepsilon_{\bm{k}} - \min \varepsilon_{\bm{k}})$ and $\varepsilon_{m,\bm{k}}{=}(1/\Delta S)\sum_{s=0}^{m}c_{s,\bm{k}}d_{s}$ (blues crosses). Blue circles and squares denote the renormalized Fock energy $\tilde{\varepsilon}_{F,\bm{k}}$ and Hartree energy $\tilde{\varepsilon}_{H,\bm{k}}$, respectively. Red solid and dotted lines denote $\text{Tr}g_{\bm{k}}$ and its various linear transformations, respectively.
(c) and (d) are similar to (a) and (b) except for $v_{1}$ interaction.
} 
\end{figure}

For $\bm{k}$-space short-range interactions, the hole dispersion is dominated by the Fock energy $\varepsilon_{F,\bm{k}}$, because the Hartree energy $\varepsilon_{H,\bm{k}}$ only involves processes with large momentum transfer and is small. For a general  $\varepsilon_{F,\bm{k}}$, we can decompose the form factor $\vert\langle u_{\bm{k}}\vert u_{\bm{k}+\bm{q}}\rangle\vert^{2}$ and the interaction $v_{\bm{q}}$ in $\varepsilon_{F,\bm{k}}$ with a set of orthonormal basis $U_{m}(a\bm{q})$ as follows
\begin{align}
\vert\langle u_{\bm{k}}\vert u_{\bm{k}+\bm{q}}\rangle\vert^{2} & =\sum_{m}c_{m,\bm{k}}(a)U_{m}(a\bm{q}),\label{eq:FF_decomposition}\\
v_{\bm{q}} & =\sum_{m}d_{m}(a)U_{m}(a\bm{q}),\label{eq:Interaction_decomposition}
\end{align}
where $c_{m,\bm{k}}(a)$ and $d_{m,\bm{k}}(a)$ are expansion coefficients, $U_{m}(a\bm{q})$ satisfies $\delta_{m,n}=\int d^{2}\bm{q}\,U_{m}(a\bm{q})U_{n}(a\bm{q})$, and $a$ is a tunable length scale. By combining Eq.$\,$(\ref{eq:FF_decomposition}), (\ref{eq:Interaction_decomposition}) and (\ref{eq:First_term}), one can obtain the Fock energy
\begin{align}
\varepsilon_{F,\bm{k}} & =\frac{1}{\Delta S}\sum_{m}c_{m,\bm{k}}(a)d_{m}(a),\label{eq:Energy_decomposition}
\end{align}
where $\Delta S=\vert\bm{b}_{1}\times\bm{b}_{2}\vert/N_{s}$ and $N_{s}$ is the number of unit cells. Even though the choice of orthonormal basis $U_{m}(a\bm{q})$ is arbitrary, one should try to pick a ``good'' $U_{m}(a\bm{q})$ such that $c_{m,\bm{k}}(a)d_{m}(a)$ decay fast with $m$ and the first few terms dominate. On the other hand, one can derive the expression for the quantum metric tensor from Eq.$\,$ (\ref{eq:FF_decomposition}) by Taylor expanding both sides to the second order of $\bm{q}$ and the result is 
\begin{align}
\text{Tr}g_{\bm{k}} & =-\frac{1}{2}\sum_{m}c_{m,\bm{k}}(a) \left. \Delta U_{m}(a\bm{q}) \right|_{\bm{q}=0},\label{eq:Trg_decomposition}
\end{align}
where $\Delta\equiv\partial_{q_{x}}^{2}+\partial_{q_{y}}^{2}$. Note that there is no need for generalized anisotropic basis here, because anisotropic terms have no contribution to the energy or $\text{Tr}g_{\bm{k}}$$\,$\citep{yang_generalized_2017}. In the following, we will analyze the relationship of $\varepsilon_{F,\bm{k}}$ and $\text{Tr}g_{\bm{k}}$ based on Eq.$\,$(\ref{eq:Energy_decomposition}) and (\ref{eq:Trg_decomposition}). 

For ideal bands, we can choose the Laguerre polynomials $U_{m}(\bm{q}l_{B})=(l_{B}/\sqrt{\pi})L_{m}(\vert\bm{q}\vert^{2}l_{B}^{2})\exp(-\vert\bm{q}\vert^{2}l_{B}^{2}/2)$ with \textbf{$l_{B}$} being the characteristic length as the orthonormal basis. For $\bm{k}$-space short-range interactions, e.g., the Coulomb interaction, both the expansion coefficients $c_{m,\bm{k}}$ and $d_{m}$ decrease with increasing $m$ in a fluctuating way. As a result, the first term $\varepsilon_{0,\bm{k}}=\frac{1}{\Delta S}c_{0,\bm{k}}d_{0}$ plays a dominant role in $\varepsilon_{F,\bm{k}}$ as shown in Fig.$\,$\ref{fig: Microscopic_analysis} (a). Meanwhile, $c_{0,\bm{k}}$ has a negative linear dependence on $\text{Tr}g_{\bm{k}}$. As a result, the leading term $\varepsilon_{0,\bm{k}}$ can well account for the linear relationship between $\varepsilon_{F,\bm{k}}^{\text{IB}}$ and $\text{Tr}g_{\bm{k}}$ for $\bm{k}$-space short-range interactions as shown in Fig.$\,$\ref{fig: Microscopic_analysis} (b). Note that the negative linear relationship between $c_{0,\bm{k}}$ and $\text{Tr}g_{\bm{k}}$ is nontrivial because it indicates $\text{Tr}g_{\bm{k}}=(2l_{B}^{3}/\sqrt{\pi})\sum_{m}c_{m,\bm{k}}(m+1/2)$ is not dominated by the first term $(l_{B}^{3}/\sqrt{\pi})c_{0,\bm{k}}$, but by larger-$m$ terms, which have an opposite contribution to the linear relationship. This is not strange because $(m+1/2)$ increases with $m$ and higher-order terms can play a more important role when $c_{m,\bm{k}}$ don't decay fast with increasing $m$.

Different from $\bm{k}$-space short-range interactions, the expansion coefficients $d_{m}$ for $\bm{k}$-space long-range interactions fluctuate greatly. Here, we take $v_{1}$ interaction as an explicit example. Its expansion coefficients are shown in Fig.$\,$\ref{fig: Microscopic_analysis} (c). Due to the large fluctuation of $c_{m,k}d_{m}$, one has to consider about a dozen of terms $\varepsilon_{12,\bm{k}}=\frac{1}{\Delta S}\sum_{m=0}^{12}c_{m,\bm{k}}d_{m}$ to reproduce the spectrum $\varepsilon_{F,\bm{k}}^{\text{IB}}$ as shown in Fig.$\,$\ref{fig: Microscopic_analysis} (d). Even so, one still can't reproduce the whole hole dispersion $\varepsilon_{\bm{k}}^{\text{IB}}$, because $\varepsilon_{\bm{k}}^{\text{IB}}$ is not dominated by the Fock energy $\varepsilon_{F,\bm{k}}$ but unexpectedly dominated by the Hartree energy $\varepsilon_{H,\bm{k}}$ with only large momentum transfer. Interestingly, we find both $\varepsilon_{F,\bm{k}}$ and $\varepsilon_{H,\bm{k}}$ have a linear relationship with $\text{Tr}g_{\bm{k}}$ for a wide range of interactions, including $v_{C}$, $v_{1}$, and $v_{3}$. The physics related to the Fock energy is beyond our current microscopic analysis.

\textit{Engineering of PH symmetry in flat Chern bands.---}
Due to the presence of emergent hole dispersion, there can be the absence of PH symmetry for two-body interaction projected into a single band, a generic feature for any topological band with non-uniform QGT, in contrast to the LL systems. Importantly, we can quantify the amount of PH breaking by the bandwidth of the hole dispersion: the PH symmetry is broken if and only if the hole dispersion bandwidth is \emph{nonzero}. It thus depends on the details of interaction. It is interesting to note that for Chern bands where the hole dispersion is well determined by the trace of the quantum metric for a large family of interactions, PH symmetric interactions can be readily engineered. This is because for any two-body interaction $v^{\left(\alpha\right)}$ that gives the hole dispersion $\varepsilon_{\bm{k}}\approx \alpha \text{Tr} g_{\bm{k}}$, the interaction $\alpha_2 v^{\left(\alpha_1\right)}-\alpha_1 v^{\left(\alpha_2\right)}$ will give an almost flat hole dispersion with emergent PH symmetry, despite the non-uniform QGT.

\begin{figure}[t] 
\includegraphics[width=0.98\columnwidth]{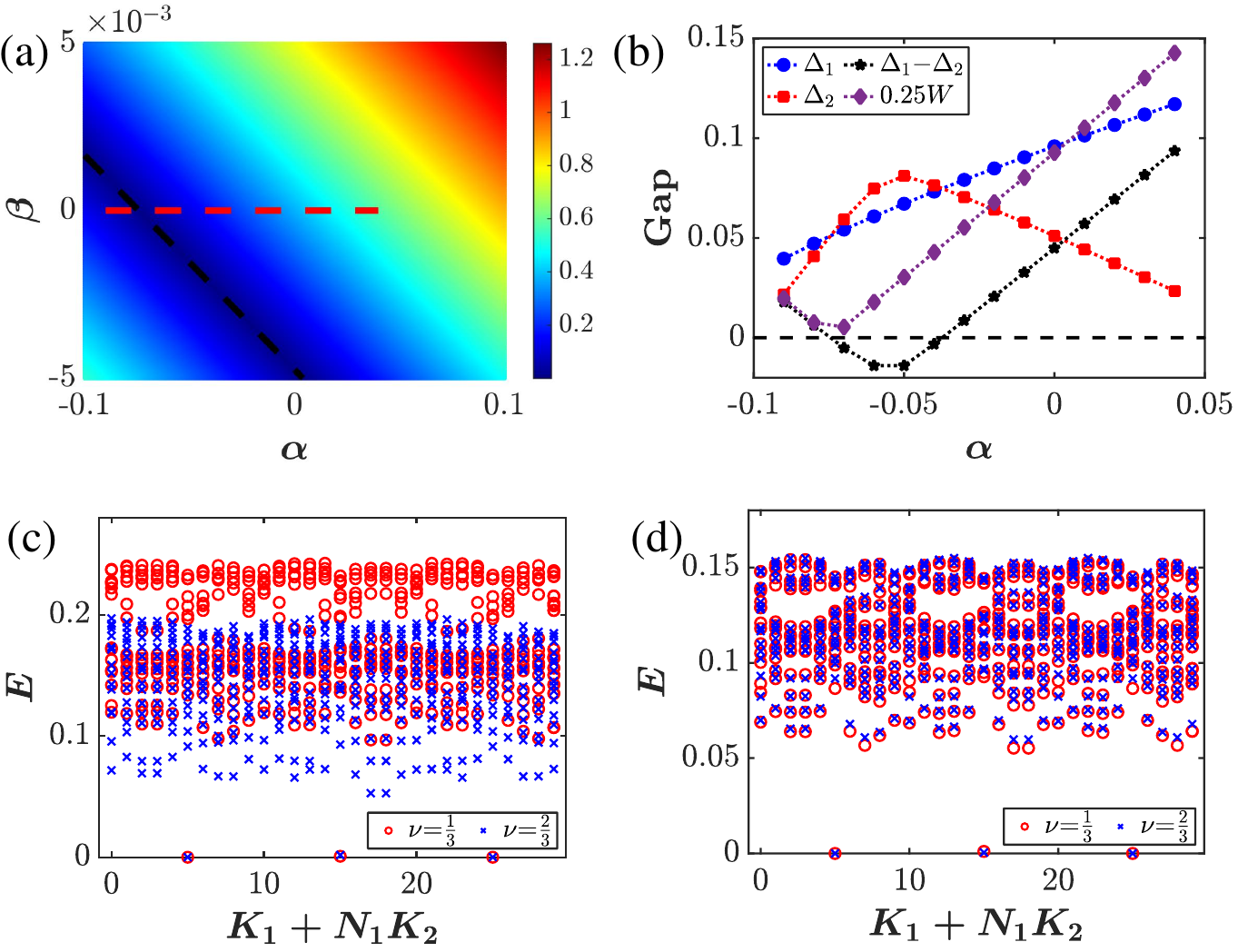}
\caption{\label{fig: Interacting_phase}
(a) Single-hole bandwidth of cTBG at the first magnetic angle for the tunable interaction $v_{C}{+}\alpha v_{1}{+}\beta v_{3}$. The system size is $N_{s}{=}5{\times} 6$. 
(b) $\Delta_{1}$ (blue circles) and $\Delta_{2}$ (red squares) denote the gaps at $\nu{=}1/3$ and $2/3$, respectively. Here, we define the gap as the difference of the highest-energy quasi-degeneracy ground state and the lowest-energy excitation state. The parameters are taken along the red dash line in (a) with $\beta{=}0$. Black pepentagrams denote the gap difference $\Delta_{1}{-}\Delta_{2}$. Purple diamonds denote one quarter of single-hole bandwidth $W$.
(c) Energy spectrum for the interaction $v_{C}$  at filling $\nu{=}1/3$ (red circles) and $2/3$ (blue crosses). Only $15$ lowest energies are plotted in each momentum sector. Integers $K_{1}$ and $K_{2}$ for $\nu{=}1/3$ are the many-body momentum indices along $\bm{b_{1}}$ and $\bm{b_{2}}$ directions, respectively$\,$\citep{wang_exact_2021}. $K_{1}$ and $K_{2}$ for $\nu{=}2/3$ are the PH conjugate of the real momentum indices.
(d) is similar to (c) except for the interaction $v_{C}{-}0.07v_{1}$. 
} 
\end{figure}

We will now take the cTBG as an explicit example. Since the linear relationship between $\varepsilon_{\bm{k}}$ and $\text{Tr} g_{\bm{k}}$ is  fairly accurate for $v_{C},v_{1},v_{3}$ (see Fig.$\,$\ref{fig: Hole_energy_vs_QGT} (a)), we can employ the following tunable interaction to engineer a PH symmetric interaction:
\begin{equation}
v_{\bm{q}}=v_{C}+\alpha v_{1}+\beta v_{3},
\end{equation}
where $\alpha$ and $\beta$ are tunable parameters. Indeed, by properly tuning $\alpha$ and $\beta$, one can get nearly flat hole bands as shown in Fig.$\,$\ref{fig: Interacting_phase} (a) (black line). To explicitly analyze the effect of the hole bandwidth on the properties of the system at fractional fillings, we take the region marked with the red dash line with $\alpha\in(-0.09,0.04)$ and $\beta=0$ as an explicit example, where both two ground states at $\nu=1/3$ and $2/3$ are incompressible FCI states. For the Coulomb interaction, the hole bandwidth is not small so that there is no PH symmetry, also reflected in the many-body energy spectrum of $\nu=1/3$ and $2/3$ as shown in Fig.$\,$\ref{fig: Interacting_phase} (c). In contrast, by introducing a small $-0.07v_{1}$ interaction, the single-hole bandwidth is greatly reduced and the PH symmetry is well restored as shown in Fig.$\,$\ref{fig: Interacting_phase} (d). Away from the hole bandwidth minimum $\alpha\approx-0.07$, the gap difference between $\nu=1/3$ and $2/3$ will usually increase except for a small region $\alpha\in(-0.07,-0.05)$, where the gap difference is also small. If we continuously increase $\alpha>-0.05$, the single-hole bandwidth will continue to increase, the gap at $\nu=2/3$ will gradually close and the system experiences a phase transition to a compressible Fermi liquid state$\,$\citep{liu2024broken}. In contrast, the topological gap at $\nu=1/3$ is always very robust.

Similarly, we can also engineer PH symmetric interactions in real materials, as long as the hole dispersions for any two kinds of interactions are proportional to each other to a good approximation. For example,  the hole dispersions of $t\textrm{WSe}_{2}$  for $v_{C}$ and $v_{1}$ are well proportional to each other as shown in Fig.$\,$\ref{fig: Hole_energy_vs_QGT} (d). Based on this result, a PH symmetric interaction $v_{C}-0.007v_{1}$ can be engineered. On the other hand, to engineer PH symmetric interactions in more complex systems carrying an electron dispersion and greatly deviating from the trace condition like $t\textrm{MoTe}_{2}$, one not only needs to consider the hole dispersion but also the electron kinetic energy and their interplay. These are beyond this work and left as further works. 

\textit{Summary and outlook.---}
We have revealed the connection between the single-hole dispersion and the metric tensor in flat Chern bands and shown how to exploit this to engineer PH symmetry in a Chern band with fluctuating QGT. The generalization to the quasihole excitation at fractional fillings and the many-body QGT could also be interesting. Over the past few years, there has been a heated debate on the many-body QGT in strongly correlated systems. For example, two different  kinds of definitions for the Berry phase in the composite Fermion liquid have been proposed in Refs.$\,$\citep{geraedts2018,wang2019lattice,ji_berry_2020} and they give different results even for the same kinds of wave functions. It is promising that one can get some understanding of the many-body QGT from the perspective of the single quasihole energy dispersion. 

Recently, it is also found that the metric tensor plays an important role in flat-band superconductors$\,$\citep{tian2023evidence,chen_ginzburg-landau_2024,hu2023anomalous}, where the superconducting coherence length is proposed to be dominantly determined by the metric tensor. This is another interesting case of strong interaction effect dominated by the single particle quantum geometry, with very strong numerical evidence though without rigorous analytic understanding beyond the long wavelength limit. It is thus useful to generalize our microscopic analysis to these cases as well, with potential fundamental connections between the two systems.

Besides analyzing the system's properties from the $\bm{k}$-space perspective as done in this work, we can also do it  from the real-space perspective. As we know, ideal bands can be mapped to LLs in a curved space$\,$\citep{Estienne_Curve_2023}. Thus, it is reasonable to expect that there exist some underlying connections between the properties of holes and the geometry in real space. Besides, long-range $\bm{k}$-space interactions actually correspond to short-range real-space interactions, in particular for $v_{1}(\bm{r})=\delta^{\prime\prime}(\bm{r})$, which is the shortest one in real space for fermions. Thus, the universal hole dispersion for a wide range of interactions in $\bm{k}$-space might originate from the short-distance physics in real space. It is promising that we can get a deeper understanding of the observations in this work by the further exploration in real space. 

\begin{acknowledgments}
We thank Jie Wang, Zi-Yang Meng, Kam Tuen Law, Zhao Liu, Jin-Xin Hu, Yuzhu Wang for valuable discussions. This work is supported by the National Research Foundation, Singapore under the NRF fellowship award (NRF-NRFF12-2020-005).
\end{acknowledgments}

\bibliographystyle{apsrev4-1}
\bibliography{FCI}

\end{document}